\documentclass[sigconf]{acmart}

\usepackage{booktabs} 
\usepackage{graphicx}
\usepackage{amssymb}
\usepackage{amsmath}
\graphicspath{{/Users/caitlingray/Documents/InformationDiffusion/Papers/figures/}}

\setcopyright{rightsretained}
\newlength{\figwidth}
\begin{document}
\newcommand{\ER}{Erd{\"o}s-R{\'e}nyi }
\newcommand{\BA}{Barab{\'a}si-Albert }

\copyrightyear{2018}
\acmYear{2018} 
\setcopyright{iw3c2w3}
\acmConference[WWW '18 Companion]{The 2018 Web Conference Companion}{April 23--27, 2018}{Lyon, France}
\acmBooktitle{WWW '18 Companion: The 2018 Web Conference Companion, April 23--27, 2018, Lyon, France}
\acmPrice{}
\acmDOI{10.1145/3184558.3191590}
\acmISBN{978-1-4503-5640-4/18/04}
\fancyhead{}

\title{Super-blockers and the Effect of Network Structure on Information Cascades}

\author{Caitlin Gray}
\affiliation{%
	\institution{The University of Adelaide}
}
\affiliation{ARC Centre of Excellence for Mathematical and Statistical Frontiers}
\email{caitlin.gray@adelaide.edu.au}

\author{Lewis Mitchell}
\affiliation{%
\institution{The University of Adelaide}
}
\affiliation{ARC Centre of Excellence for Mathematical and Statistical Frontiers}
\email{lewis.mitchell@adelaide.edu.au}

\author{Matthew Roughan}
\affiliation{%
  \institution{The University of Adelaide}
}
\affiliation{ARC Centre of Excellence for Mathematical and Statistical Frontiers}
\email{matthew.roughan@adelaide.edu.au}

\def\change{\textcolor{blue}}
\def\lewis{\textcolor{orange}}
\def\matt{\textcolor{blue}}
\def\question{\textcolor{red}}
\newcommand{\lm}[1]{{\color{orange}[[LM: {#1}]]}}

\renewcommand{\shortauthors}{Gray \emph{et al.}}

\begin{abstract}
Modelling information cascades over online social networks is important in fields from marketing to civil unrest prediction,
however the underlying network structure strongly affects the probability and nature of such cascades.
Even with simple cascade dynamics the probability of large cascades are almost entirely dictated by network properties,
with well-known networks such as Erdos-Renyi and Barabasi-Albert producing wildly different cascades from the same model. 
Indeed, the notion of `superspreaders' has arisen to describe highly influential nodes promoting global cascades in a social network. 
Here we use a simple model of global cascades to show that the presence of locality in the network increases the probability of a global cascade due to the increased vulnerability of connecting nodes. 
Rather than `super-spreaders', we find that the presence of these highly connected `super-blockers' in heavy-tailed networks in fact reduces the probability of global cascades, 
while promoting information spread when targeted as the initial spreader.
\end{abstract}

%
%

\ccsdesc[500]{Networks~Network structure}
\ccsdesc[300]{Human-centered computing~Empirical studies in collaborative and social computing}

\keywords{Network Structure; Information diffusion; Cascades}

\maketitle

\section{Introduction}
The movement of information through social networks is a phenomenon observed online in the spread of ideas, pictures and products. Such movement is important in numerous fields, from online marketing to the prediction of civil unrest events \cite{Cadena-15}.  This work addresses how the properties of the underlying structure of the social network, such as locality, affect the flow of information.

Online social networks, such as Facebook and Twitter, can provide large volumes of network data; however, these social networks can consequently be computationally expensive to work on. 
More importantly, rate limits (Twitter) or private data (Facebook) often make it infeasible to collect even a small portion of the network structure. 
It is also difficult to distinguish the effect that properties of interest may have on information flow without comparison networks. 
Therefore, random graphs are an essential tool for studying information cascades,
allowing for the controlled variation of the network properties of interest.

It is often noted that when faced with a decision to change their behaviour, for example, to adopt a new product or share a post on a social network, people display inertia. 
For an individual to promote social spread of an idea, multiple exposures are often required. 
This may occur in situations when individuals do not have enough information to make a decision, or individuals do not wish to share online content unless many neighbours already are. 
This motivates the use of threshold models in information cascade modelling.   
Watts presents a threshold model of information cascades \cite{Watts-02} that provides a platform to explore the effect of the underlying graph structure on cascades.

Here, we investigate the effect of changing certain properties of the underlying network structure on the probability and frequency of information cascades.  We use the Watts model of information cascades as the basis for our investigation, and model the network using three random network models with varying parameters.

There is an extensive literature on the effect of complex network structure on epidemiological models \cite{May-01, Moreno-02}.  A major theme arising in these papers is the presence of `super-spreaders' \cite{Lloyd-Smith-05}: nodes of high degree that can infect many people. 
These super-spreaders are intuitive in the epidemiological context,
due to the nature of the transmission dynamics governing disease spread.
In contrast, as we will show in this paper, the multiple exposures required prior to activation mean that information flow is in fact inhibited by the same type of high-degree nodes,
which we will call `super-blockers'.

The location of individuals in a social network is often a factor in the initiation and maintenance of connections, so social networks often display dependence on proximity. We explore the effect of network locality, and resulting clustering, to show that increased dependence on proximity increases the frequency of large cascades.

Our main results concerning the impact of network structure are described in \autoref{results}, once we have provided a precise description of the models and methods to be used.

\section{Background}
Random network models provide a framework to investigate the effect of network properties on information cascades.  Here we present the mathematical formulations of three basic random networks that show different levels of locality and clustering, on which we will model information cascades.
\subsection{Network Models}
A network or graph $\langle V,E \rangle$ is a collection of $n$ nodes, connected by $e$ edges, where $n =\vert V \vert$ and $e =\vert E \vert$. Many properties are defined to describe the structure of the network, see \cite{Networks}.  The \emph{degree} of the $i$th node, $z_i$, is the number of edges incident to it.  In social networks these can represent social connections or friends/acquaintances. The \emph{average degree} of a network is denoted $z$. 


Social networks commonly have spatial structure as people in close contact are more likely to be friends with each other \cite{Scellato-11}. 
Online social networks also display this tendency, albeit slightly less strongly than in other social networks, as it is easier to maintain longer connections online \cite{Ellison-06}. 
The Waxman graph \cite{Waxman-88} is a spatially embedded network commonly used in the topology of physical networks, and reduces the probability of long links. The ratio of short to long links can be tuned for desired properties, such as clustering and betweenness.



Many extensions to the Waxman graph have been proposed, \emph{e.g.} \cite{Naldi-05}. However, in some later formulations the notation has become confused. An alternative parametrisation is used here \cite{Roughan-15}: the probability of attachment between two nodes $u$ and $v$ separated by distance $d$ is
\begin{equation}
\label{waxman_eq}
P(u,v)=q e^{-sd},
\end{equation}
for $q\in (0,1]$, $s\ge0$. The parameter $s$ controls the extent to which spatial structure is incorporated into the graph. The $q$ value is the thinning of edges in the graph and often $q$ is restricted to $(0,1]$ \cite{Roughan-15}.  Larger $s$ values decrease the likelihood of longer links, and increase the clustering.

Here, we use the term \emph{locality} to describe the extent to which a network's links dependence on distance. This means that nodes that are distant are less likely to be connected. In Waxman networks, higher $s$ values show more locality.  Although related, this is different to the resulting \emph{clustering} in the network. The \emph{clustering} of a network is the extent to which the friends of an individual $i$ are also friends with each other. This is routinely observed in social networks with `cliques' or \emph{clusters} seen in both real world and on-line networks.



When $s = 0$ in (\ref{waxman_eq}) the Waxman graph becomes the well-known Erd{\"o}s-R{\'e}nyi (ER) graph $G(n,q)$ with $n$ nodes and probability of attachment $q$ \cite{ER-59}. 
This mathematically tractable graph has been studied extensively\cite{Bender,Bollobas-80,Molloy-95,NewmanStrogatz}.  
Notably, the $s=0$ construction produces graphs that do not exhibit clustering or highly connected nodes.


At the other extreme with respect to clustering is the \BA (BA) graph \cite{Barabasi-99},
which can more realistically describe some real world networks such as the World Wide Web (WWW) or some social networks. 
It was motivated by the observation that many real networks are connected by a power-law degree distribution, driven by incoming nodes preferentially attaching to highly connected nodes.  
While there is currently debate about how frequently these `scale-free' networks occur \cite{Broido-18}, at the very least it provides a contrast model on which to test cascade dynamics.  

As the network grows each node has fixed integer $m$ initial connections upon entering the graph, so the average node degree is $z = 2m$ \cite{Networks}. Both growth and preferential attachment are sufficient to produce power-law distributions of connectivities. Extensions of the BA graph consider altering the model to use non-linear attachment probability and adapting the growth heuristic to include node and edge removal~\cite{Boccaletti-06}.


The Price random graph generalises the \BA graph by using a Poisson random value of initial connections, rather than a fixed value \cite{Price-76}. That is, each new social network user will not have the same number of initial friends. 

Finally, it is important to note that when simulating random graphs some parameters or properties are fixed but the actual connections of the graph differ each time they are created. A given graph is a single \emph{realisation} of a statistical ensemble of all possible combinations of connections \cite{Boccaletti-06}.

\subsection{Information Cascades}
Watts presents a simple model of global cascades \cite{Watts-02} to model the flow of information on random networks that incorporates a threshold function to model binary decision making.

The model starts with a network of $n$ nodes, initially in an inactive state, and a shock is introduced to the system, \emph{i.e.} one node is made active, to initialise the cascade. The state of node $i$ at time-step $t$ is given by
\begin{equation}
s_i^t = 
\begin{cases}
1, & \text{if active},\\
0, & \text{otherwise}.
\end{cases}
\end{equation}
The population then evolves at successive time steps in which all nodes simultaneously update their state according to the threshold rule:
\begin{equation}
s_i^{t+1} = 
\begin{cases}
1, & \text{if } s_i^t = 1 \text{ or } \sum\limits_{j \in N(i)} s_j^t \text{   }> \phi_i z_i , \\
0, & \text{otherwise.}
\end{cases}
\end{equation}

Where $\phi_i$ is the threshold of a node, taken from $f(\phi)$, where $f$ is an arbitrary distribution on (0,1], and $N(i)$ is the set of neighbours of $i$. Each node observes the current states of its $k$ neighbours and becomes \emph{active} if at least a proportion $\phi_i$ of its neighbours are active.
Once a vertex has become active, it remains active for the duration of the cascade, and the process terminates when no further changes are observed.
The \emph{stability} of a node is a measure of how susceptible the node is to outside influence, and is an important factor in the percolation of information through a network. 

The stability of the node is given by $\kappa_i=\lceil \phi_i z_i \rceil$, and is the number of active neighbours required before the node will be activated. This shows that nodes with more neighbours will be less influenced by the activation of an individual neighbour. A node in the network is defined as vulnerable if it has $\kappa<1$; that is, if $z_i <\lfloor{1/\phi_i}\rfloor$ and is activated by a single active neighbour.

Epidemiological models are commonly used to model disease outbreaks on networks, such as the BA network.  This leads to the notion of `super-spreaders': highly connected nodes at the centre of the graph that act to accelerate disease transfer. We will focus on the difference between super-spreaders in the disease context and stable nodes for information transfer, and argue for the existence of `super-blockers'.

\section{Methods}
Our goal is to understand the cascade behaviour on different types of random graphs. To that end we primarily use simulations, as, although Watts' model is analytically tractable on ER graphs, the analysis techniques use properties of the ER random graphs that do not extend to other random graphs of interest.

We investigate the effect of changing the locality structure on information cascades. This is achieved by changing the $s$ parameter for the Waxman network controlling the dependence on distance between nodes.  

Super-blockers are highly connected nodes that require multiple exposures to propagate information. The \BA and Price networks are well known for the presence of these hubs and are used to investigate the role they play in information diffusion.

The NetworkX package (version 1.11) in Python (version 2.7) \cite{python} was designed to create, manipulate and analyse complex networks and is used here. The \BA random graph was created using the inbuilt NetworkX function, and the Price network was created by altering the BA algorithm to include the random variation of initial connections. 

The inbuilt Waxman generator was used to create the random graph after determining the parameter $q$ from the required $s$ and $z$ values using the equation derived in \cite{Roughan-15}:
\begin{equation}
\label{q_equation}
 q = \frac{z}{(n-1) G(s)}.
\end{equation}
Where $G(s)$ is the Laplace transform of the probability density function, $g(t)$, for the line-picking problem.

Watts' cascade model was implemented on 10 realisations of these graphs with $n = 10, 000$, and initial shocks containing a single randomly selected node. All simulations were implemented with $k = 1,000$ random initial shocks per network. The process terminates when no new nodes are activated in a single step, and the size of the cascade is recorded. As in Watts' work, the thresholds are given by a delta function $f(\phi)=\delta(\phi-\phi^*)$ where $\phi^*$ is a constant. 

In the following experiments $\phi^*=0.18$ is used for consistency with Watts' work \cite{Watts-02}, although similar qualitative results hold for different $\phi^*$ values. Stable nodes in these networks have $z_i>4$, while vulnerable nodes have $z_i\le4$.

\section{Global Cascades}
A global cascade, ideally, is one that propagates through the entire network until all nodes are active and there is \emph{global} adoption of the idea. However, this definition is inappropriate in many situations due to poor connectivity and the unrealistic assumption that everyone in a network must participate for a notion to be considered global in real world events. Unfortunately in the literature the definition of a global cascade varies. One widely used definition is that a global cascade is a cascade that occupies a given fraction of a network~\cite{Watts-02}; however, the fraction varies or is unstated.

Here, the following two definitions were considered:
\begin{enumerate}
\item A global cascade occurs when the largest possible cascade occurs (\emph{i.e.}, it covers the largest connected component).
\item A global cascade occurs when greater than a proportion $b$ of the network is activated in a cascade.
\end{enumerate}

A global cascade can be defined as the maximum possible cascade size for the network being considered. The connectivity of the network will determine the size of the giant component, and hence the maximum size cascade possible. This is found here empirically as the maximum proportion observed in a large number of simulations, and defined as a global cascade for that network. This definition is more appropriate for networks that are not necessarily completely connected. However, this does not count cascades only slightly smaller than the maximum.

The second definition can be customised to use any value $b$.  Using $b=0.99$ gives the intuitive idea that a global cascade is a cascade in which the entire network is activated, while allowing for a small fraction to be inactive. Lower values of $b$ are acceptable as in reality few trends are ever adopted by 100\% of potential participants. However, using this value of $b$ does not account for networks that are not fully connected.


In the Watts model the distribution of cascade sizes is generally bimodal for $z>1$. That is, cascades are either very small or very large \cite{PayneDodds-09, Watts-02}. For bimodal cascades there will be a value $b$, much less than 1, above which the maximum cascades occur. Some studies \cite{WattsDodds-07} use $b=0.01$. 

In our results, it was found that cascades were generally bimodal and $b=0.1$ will encompass all cascades that can be considered global. Watts also considered various $b$ values for robustness and found that $0.1$ was appropriate \footnote{D. Watts, Personal Communications, September 2016}.  We found that the specific value of $b$ does not significantly affect the results.

Both definitions of global cascades were tested in each context. In cases that are not bimodal a global cascade is determined empirically by the largest cascade observed on the graph.  However, we find that the Waxman, \BA and undirected Price graphs give rise to bimodal cascades and so global cascades were defined as greater than 10\% of the network, similarly to Watts.

\section{Results and Discussion}
\label{results}
\subsection{The effect of locality in graph structure on cascades}
Many real world networks demonstrate locality, notably in social networks where friendship clusters arise for a number of reasons, such as proximity and shared interests. Watts introduced the simple model of global cascades for an \ER graph with effectively zero clustering, but real networks do exhibit clustering and locality. We start by investigating the effect of local structure on the probability of global cascades through the Waxman network.

As introduced above, the Waxman parametrisation used here has a parameter $s$ that determines the ratio of long to short links, and hence the clustering of the networks. Therefore, the effect of spatial structure on cascades can be investigated by changing the parameter $s$ for any $z$ values. Higher $s$ values result in networks with higher clustering. 

\begin{figure}
\centering
\includegraphics[width=8cm]{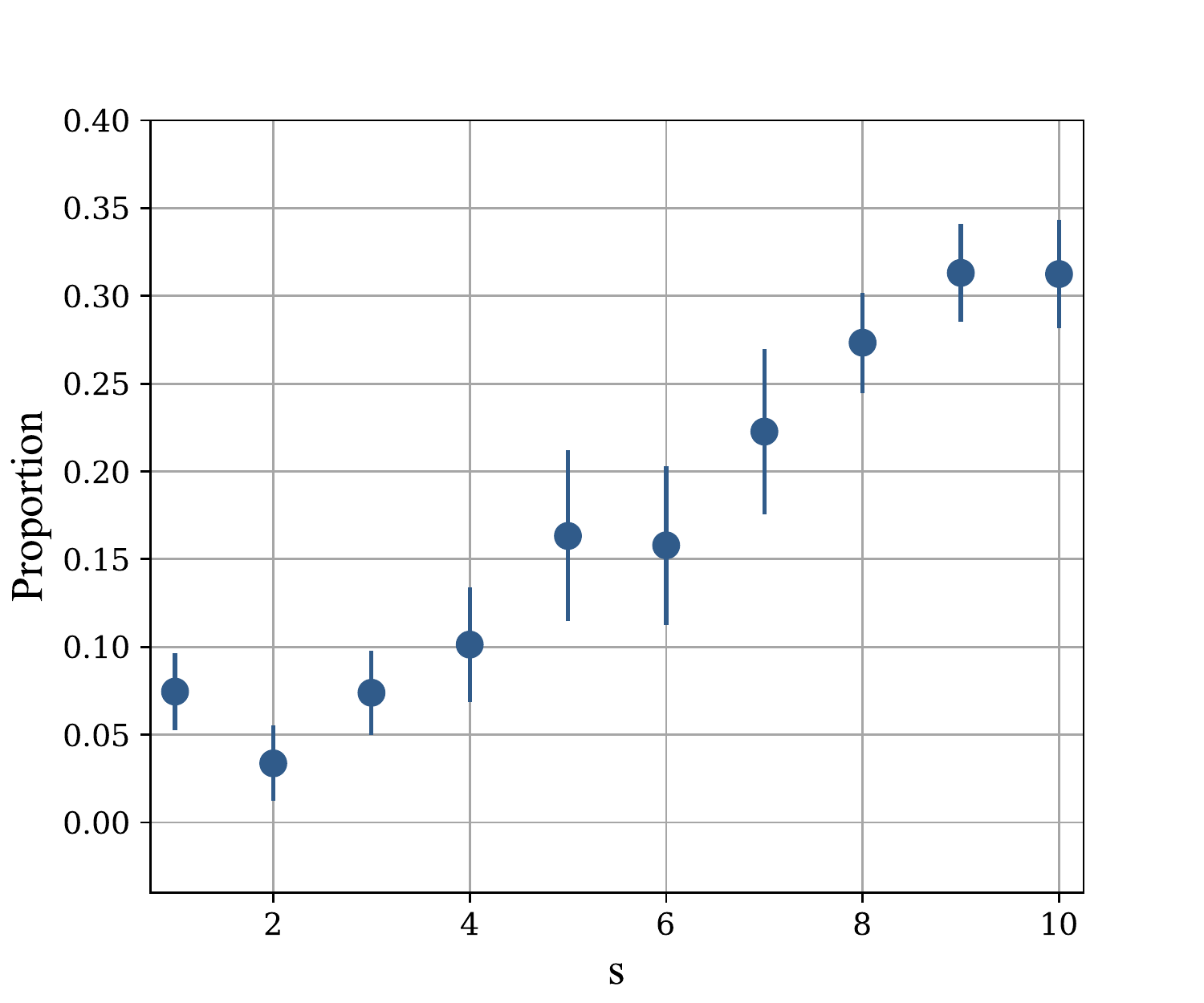}
\caption{Average frequency of global cascades on a Waxman network of $n=10,000$ nodes with $z=6$.  The average size and frequency over $k=1,000$ initial shocks on 10 realisations of the Waxman network is shown.  Larger $s$ values result in a higher probability of cascades, likely due to some nodes connecting the clusters having low degree. The error bars shown are 95\% confidence intervals.}
\label{fig:waxman_s}
\end{figure}

\autoref{fig:waxman_s} shows the results: as $s$ increases, along with clustering and locality, the probability of a global cascade is increasing.

The increase in frequency is due to the change in geometry, specifically in the degree of the `connecting nodes' of these clusters, \textit{i.e.}, nodes with long links. The connecting nodes of the graph are those with high \emph{betweenness} and are essential in the propagation of cascades.  The betweenness measures the importance of a node in a social network by considering the number of shortest paths going through it, see \cite{Boccaletti-06}. Nodes with high betweenness are considered important to the graph as they are essential to creating shorter paths between nodes.

\begin{figure}
\centering
\includegraphics[width=8cm]{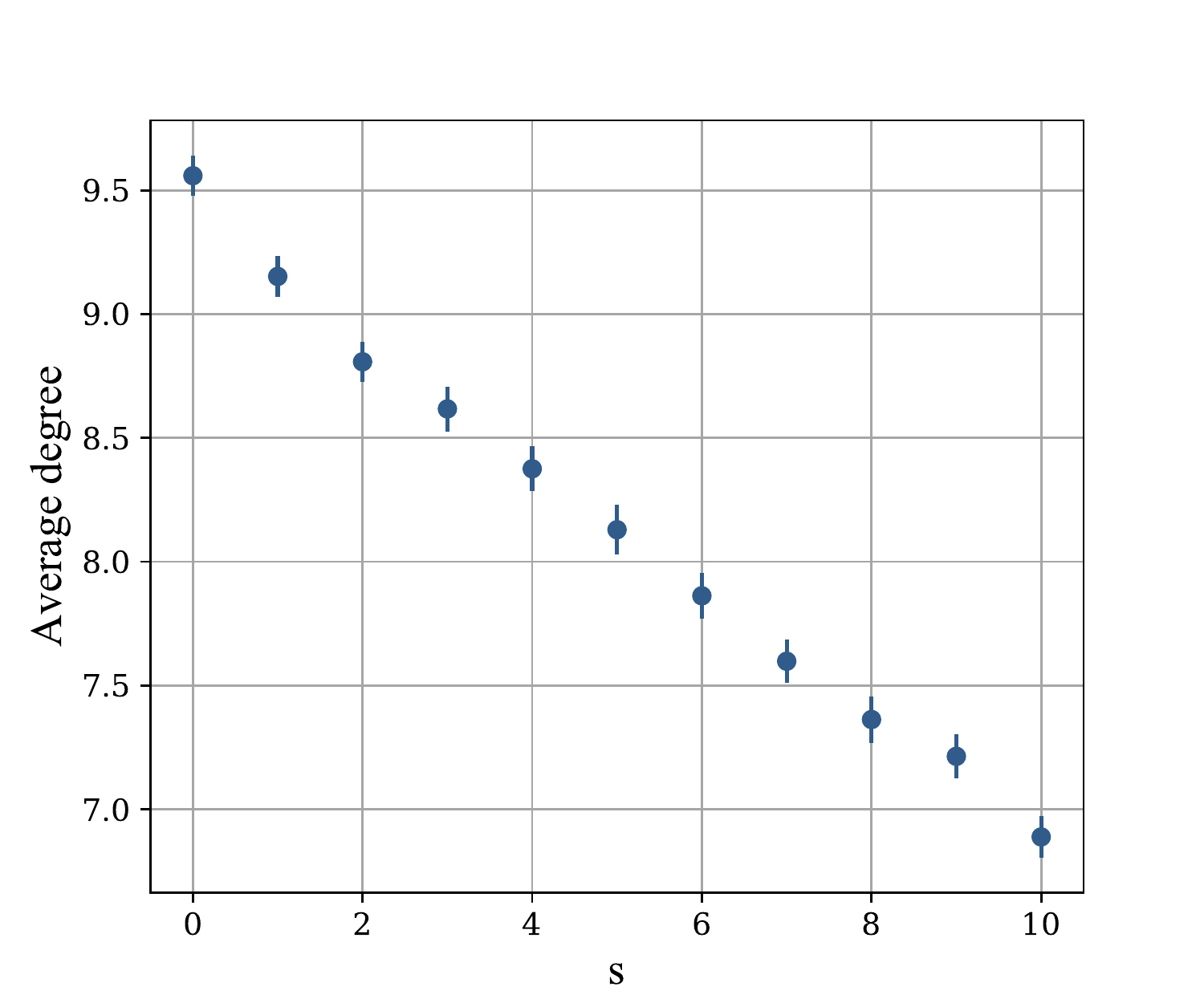}
\caption{The average degree of nodes with high betweenness (betweenness $>$ 0.03) for differing values of $s$. The average is taken over 300 network realisations for each $s$ value. Error bars show 95\% confidence intervals.}
\label{avdeg_highbet}
\end{figure}

In the ER graph, nodes with high betweenness are more likely to be connected to more nodes and so have high degree $z_i$.  This is shown in the first data point ($s=0$) of \autoref{avdeg_highbet}. This can create a `super-blocker' with higher degree and inhibits the flow of cascades.

Conversely, we show, in \autoref{avdeg_highbet} that in Waxman graphs with higher $s$, the nodes with high betweenness can have lower degree, due to the connecting structure between clusters.

\autoref{avdeg_highbet} shows the degree of nodes with high betweenness ($>0.03$). This is a measure of the stability of nodes on the shortest paths within the network.  As the clustering increases these important nodes have lower degree, and are vulnerable to activation.  Ideas are reinforced by closely connected nodes within the cluster, and connecting nodes between these clusters promote the information flow between clusters.


It is worth noting that as $s$ increases the Waxman graph becomes more clustered and more likely disconnected. However, with large networks and average degree above 2, the largest connected component will still contain $>90\%$ of the network.  This would slightly reduce frequency of large cascades as there are less seeds in the connected component. However, this small effect is dominated by the increase in clustering as described above.

\begin{figure}
\centering
\includegraphics[width=8cm]{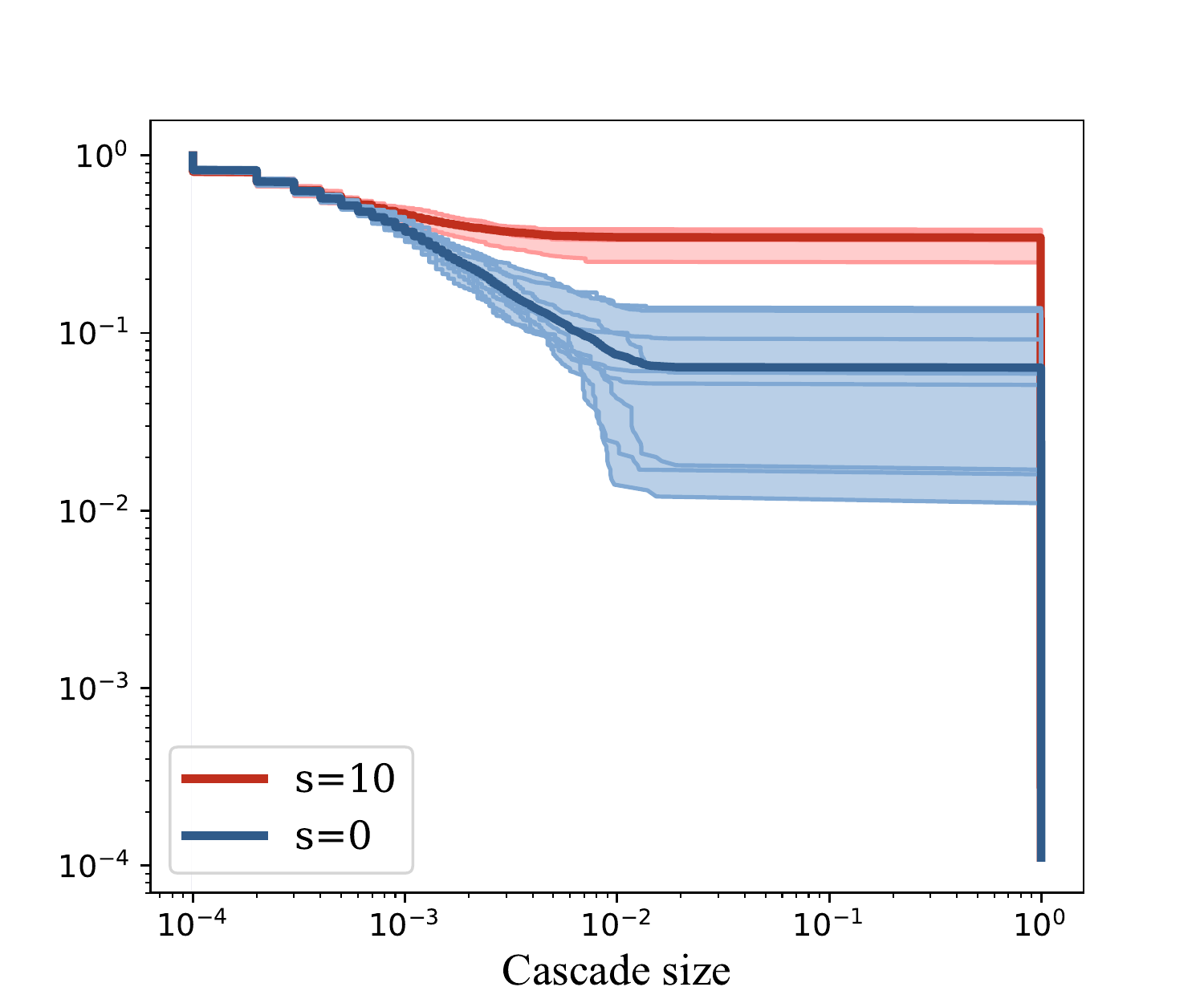}
\caption{Empirical complementary cumulative distribution of the cascade size on Waxman networks with $z=6$ using $k=1,000$ initial shocks.  The two cases $s=0$ (blue) and $s=10$ (red) are shown.}
\label{ccdf}
\end{figure}

The complementary cumulative distribution function (CCDF) of cascade size is shown in \autoref{ccdf} for the two extreme clustering scenarios: $s=0$ and $s=10$. The $s=0$ case is equivalent to the \ER network described by Watts \cite{Watts-02}. The increase in frequency of global cascades for the networks with higher clustering is evident. The clustered graph produces larger cascades with an average frequency of 31.2\% much higher than the comparative $s=0$ case of 7.46\%.
\autoref{ccdf} overlays the cascades of different realisations of the network. It is evident that the underlying graph has an impact on the distribution of cascades. Small cascades are not highly dependent on the underlying graph as all networks can facilitate small cascades from seeding either poorly connected or highly stable nodes. Conversely, the probability of large cascades is highly variable. This shows that the probability of global cascades is dependent on the specific connectivity of the underlying network in addition to the parameters of the random graph.

\subsection{The effect of average degree on cascades on Waxman networks}
Waxman graphs are represented by the two parameters $s$ and $q$. However, an arguably more meaningful parameter is average degree ($z$), determined from \autoref{q_equation}, as it determines the density of links and stability of nodes. The average degree of the network $z$ affects the overall stability of the graph and will alter the frequency of global cascades.

\begin{figure}
\centering
\includegraphics[width=8cm]{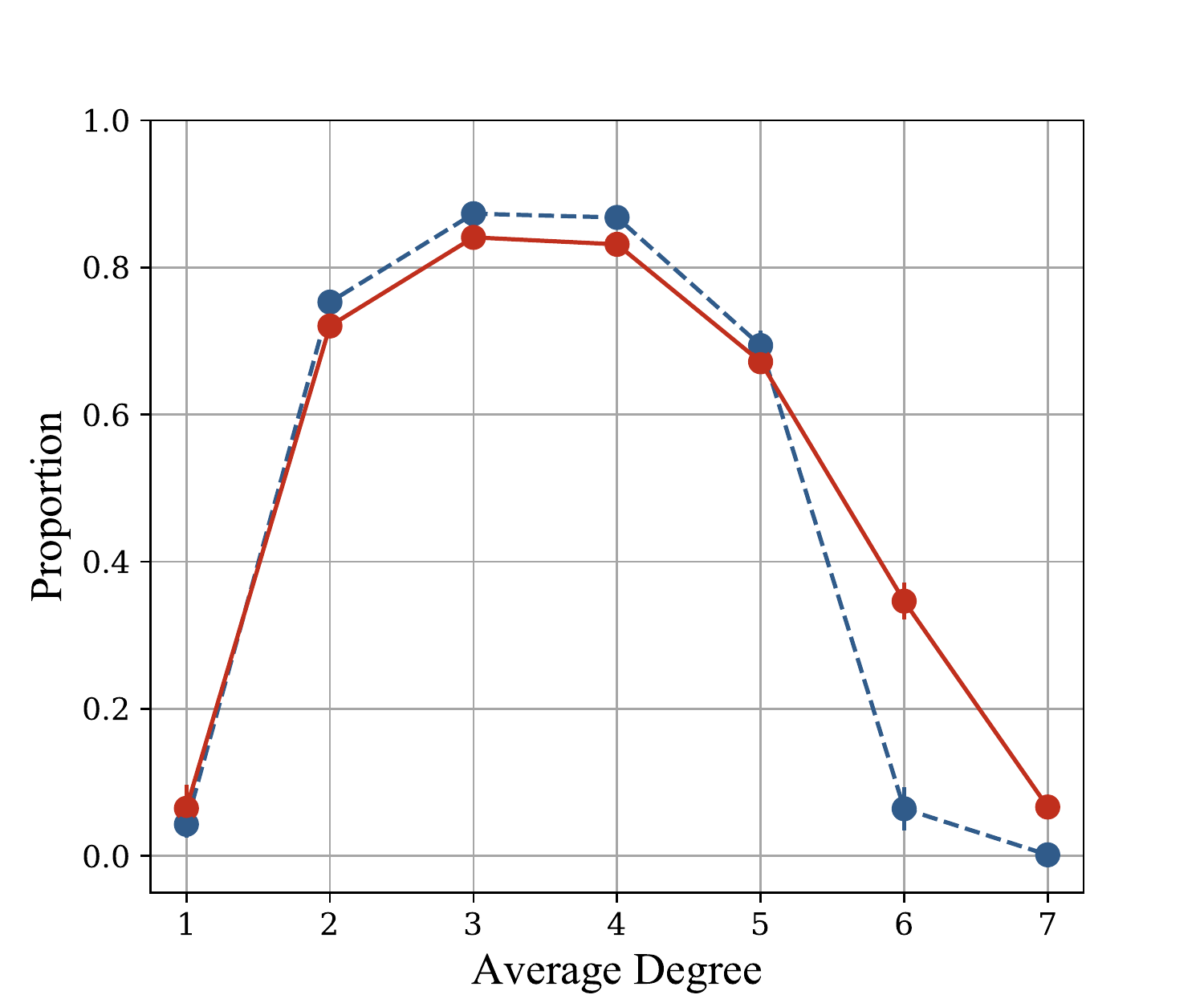}
\caption{Average frequency of global cascades on a Waxman network with $s=0$ (blue dashed) and and $s=10$ (red solid) for varying $z$. At $z=6$ where the $s=10$ case has a much higher frequency than $s=0$. Note that no global cascades occur at $z=7$ for the $s=0$ networks (as average size is zero); conversely, the high clustering at $s=10$ allow for global cascades}
\label{waxman_z}
\end{figure}


\autoref{waxman_z} also shows the frequency of global cascades for varying $z$. In the Waxman graph, as the giant component increases with $z$ the possibility of the initial shock being part of the connected component increases. For $z\le4$, most nodes still remain vulnerable, increasing the frequency of global cascades. As $z$ increases further, the stable nodes, with $z_i \geqslant 5$ in the graph become dominant, decreasing the frequency of global cascades.

\autoref{waxman_z} also shows networks with $s=10$ are slightly less susceptible to global cascades for $z\le5$; however, at $z=6$ have a higher global cascade frequency. The smaller giant component of the $s=10$ contributes to the lower frequency of global cascades for $z\le5$ as there are less shocks that will activate the giant component. The change in the relationship between the two curves from $z=4$ to $z=6$ is an interesting phenomenon caused by the emergence of locality in $s=10$ described above. In real world networks, stable nodes are common, as most individuals do not share or propagate information upon a single exposure, so networks with higher $z$ are more realistic.  It should also be noted that the high stability in $z=7$ results in no global cascades for $s=0$ case. However, the presence of clustering in the $s=10$ case allows for global cascades in these otherwise stable networks.

\subsection{Effect of degree structure on cascades}
The \BA graph \cite{Barabasi-99} is well-known for its power-law degree distribution and models the presence of high-degree nodes. In epidemiology, these nodes are often `super-spreaders'.  To investigate the effect of these hubs on information cascades, Watts model was applied to \BA and Price networks.

\begin{figure}
\centering
\includegraphics[width=8cm]{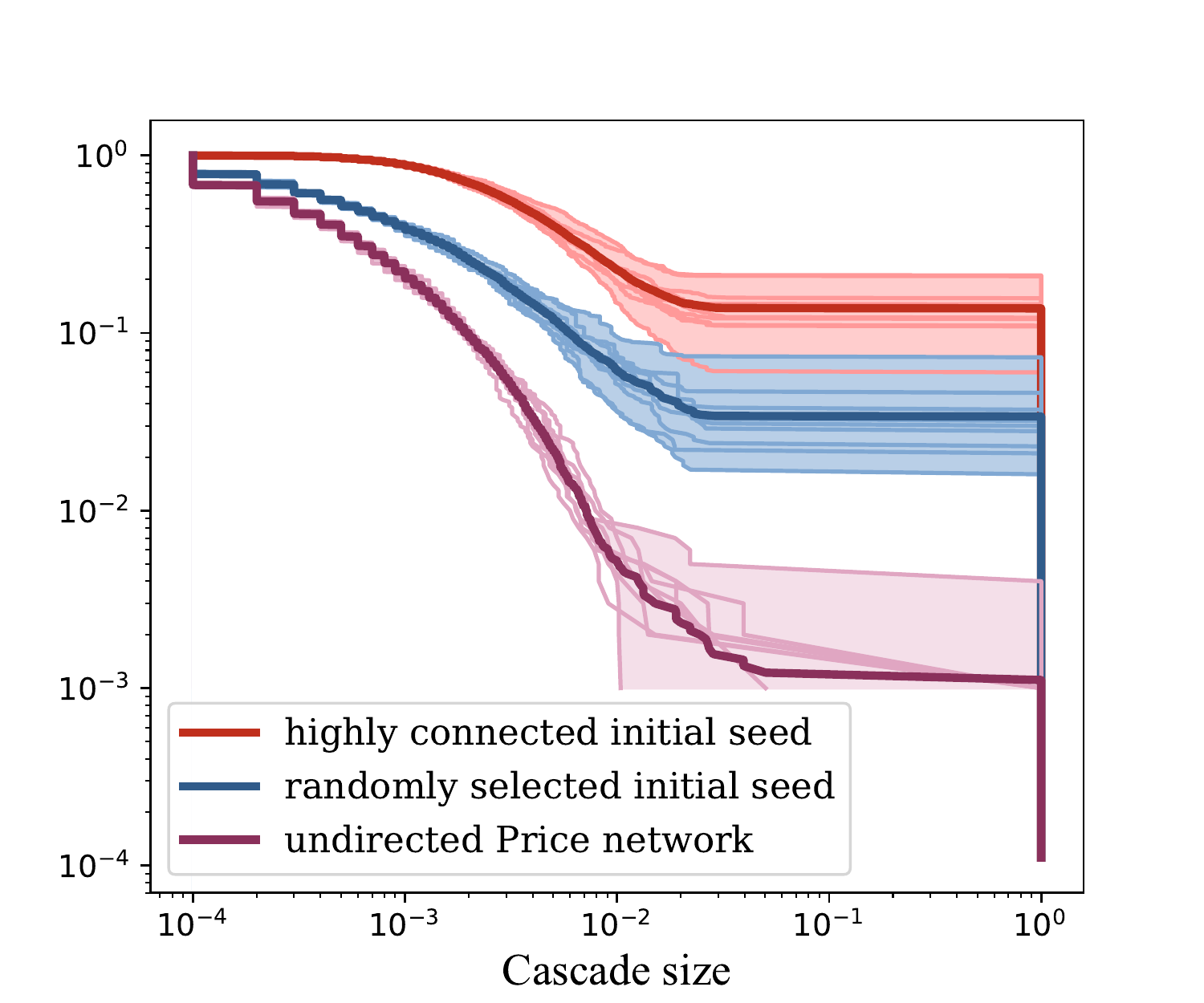}
\caption{Empirical complementary cumulative distribution of cascades size on realisations of the \BA network with $m = 3$, $n = 10, 000$.  Cascades were initialised with $k=1,000$ randomly selected nodes (blue) and only highly connected nodes (red).  The CCDF for cascade size an undirected Price network is also shown (purple).}
\label{BA_ccdf}
\end{figure}

The CCDF in \autoref{BA_ccdf} (blue) shows the cascade sizes on the BA graph. There are more cascades that do not extend past the initial shock, 21.4\% compared to 17.1\% and 18.9\% in the $s=0$ and $s=10$ Waxman graphs respectively. In the BA graph, in which nodes are by definition connected to at least $m$ other nodes, `zero cascades' are caused by the activation of nodes with small $z_i$. Nodes with small $z_i$ are added in the latter stages of the network process, and are therefore connected to high-degree neighbours. These blocking neighbours can not be activated by a single activated neighbour and so the propagation fails. This occurs often in the BA networks and is seen in social networks when an individual posts but is not shared onwards by anyone.

In information cascades, highly connected hubs with high stability act as `super-blockers', effectively partition the vulnerable nodes.  This is in contrast to the `super-spreader' phenomenon observed in simulations on networks with epidemiological models \cite{May-01}.  One possible explanation is the cognitive load required to keep track of a large number of friends/followers.  For an individual with many friends, the observation of a single person will carry less weight overall.  Therefore, a large number of exposures is required to influence the super-blockers.

Despite this, global cascades occur demonstrating that `super-blockers' can in fact aid cascade propagation if the blocker initiates the cascade. The large cascades on the BA graph occur when the initial shock hits a highly connected node, with a large number of, likely vulnerable, neighbours.


This idea can be used by marketers for promoting products on social media. They rely on the use of highly connected individuals, close to the `centre' of a networks, to advertise products and ideas. To provide evidence for this phenomenon, cascades were simulated using only initial shocks that are highly connected nodes. These `super-blockers' are stable but are expected to have neighbours with a range of node degrees. \autoref{BA_ccdf} (red) shows the cascade CCDF using only highly connected initial seeds.  It is evident that no `zero cascades' occur, compared with those of a purely random initial shock. Highly connected initial nodes increase the initial propagation through the network by activating vulnerable nodes that can then combine to finally overcome the super-blockers.

BA random graphs have a fixed integer number of connections for each node. To create more specific node degrees the Price random graph with random connections varying about $c$ was used.  \autoref{BA_ccdf} (purple) shows the CCDF of the size of cascades for Watts' cascade model applied to undirected Price networks. It is evident that very few global cascades occur compared to all other network types, despite the similarities in the BA and Price networks. This is due to the change in degree of incoming nodes. Incoming nodes of BA networks are all vulnerable and attached to by subsequent nodes with equal probability. In contrast, the incoming nodes in Price's model can have any $c_i>0$ chosen from a Poisson distribution with expected value $c$. The higher degree of some incoming nodes increases the stability of the overall graph as there are fewer vulnerable nodes in the outer region of the graph. These new nodes have a higher probability of attachment, resulting in a graph that is more spread. The effect of the increase in overall stability of the graph is seen in \autoref{BA_ccdf}.

Note that the undirected cases were used here. The directed Price model \cite{Price-76} results in a graph with direction from old nodes to new nodes. While this is useful for modelling networks like the citation graph, it is inappropriate for social networks. Social networks have an abundance of cycles and two way connections (friendships) as well as directed links (following a celebrity). In modelling information flow on directed Price networks global cascades are, as expected, extremely rare, and the distribution of cascades is a power-law, mimicking the degree distribution of the network.

\subsection{The effect of average degree on BA and Price networks}



\begin{figure}
\centering
\includegraphics[width=8cm]{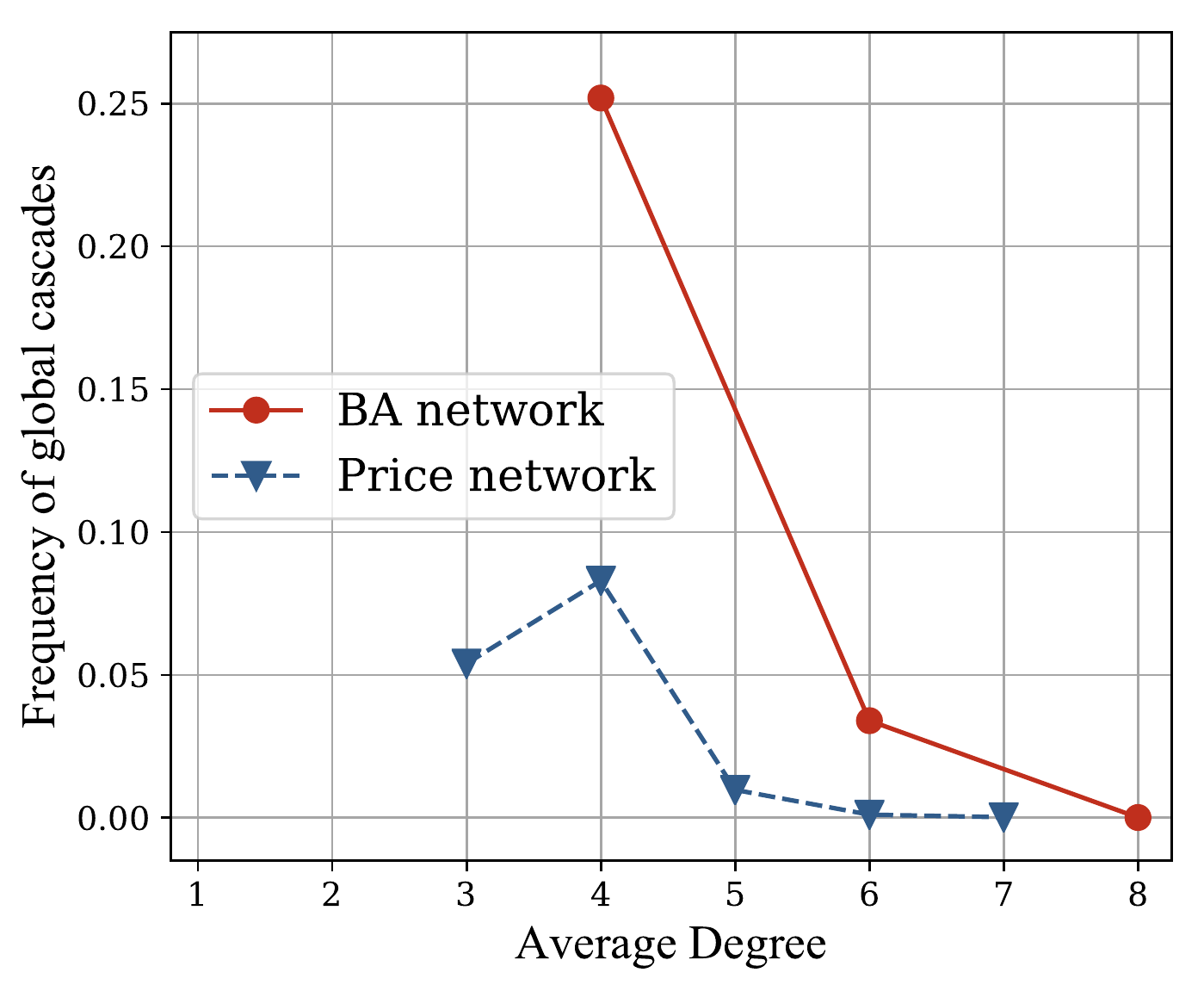}
\caption{Average frequency of global cascades for BA (red) and Price (blue) networks with $n=10,000$ nodes initialised with $k=1,000$ seeds.  Note that the BA network only allows even $z$ and we plot non-zero frequencies only.}
\label{fig:Price_freq}
\end{figure}

The average degree of the network has a distinct effect on the frequency of global cascades. \autoref{fig:Price_freq} shows the frequency of global cascades for the BA and Price networks. Cascade frequencies of zero are not shown. In both cases there are no global cascades for small $z\le2$. In these cases the graph is essentially a star, with a few central hubs and most incoming nodes connect to these hubs. This is an unrealistic representation of a social network and there are not sufficient connections to produce a global cascade. 

The maximum frequency of global cascades occurs in all graphs at $z=4$; however the frequency is significantly decreased from the $\sim$85\% observed in ER and Waxman networks. BA and Price networks have a large proportion of `super-blockers' by the preferential attachment process. This hinders the propagation in the early stages and reduces frequency of global cascades. As discussed above, the overall stability of the Price networks are higher than the BA networks resulting in a lower proportion of global cascades.

It should be noted that for the BA algorithm, $z=2m$, where $m$ is the number of initial connections of each node. Therefore, there are limited data points available; however, the use of non-integer values of $c$ in the Price networks can produce graphs with a wider range of average degree.


\section{Conclusion}
This work explores the effect of graph structure on the flow of information over a network using Watts' simple model of global cascades. Specifically, the presence of locality structure in Waxman graphs promotes the diffusion of information and enhances the frequency of global cascades. The presence of `super-blockers' in the \BA and Price networks reduce the frequency of global cascades. These results have implications for understanding and predicting information cascades in areas such as civil unrest event prediction, epidemiology and on-line marketing.  To further this work real information cascades can be used to determine the underlying network structure and determine how effectively random graphs can approximate them. The distribution of thresholds is crucial in determining the stability of the nodes. The distribution of thresholds across users and how easily they activate could be found empirically through analysing data from ego networks.

\section{Acknowledgements}
The authors acknowledge the Data to Decisions CRC (D2D CRC), the Cooperative Research Centres Programme and the ARC Center of Excellence for Mathematical and Statistical Frontiers (ACEMS) for funding this research. This research is supported by an Australian Government Research Training Program (RTP) Scholarship.


\bibliographystyle{ACM-Reference-Format}
\bibliography{Superblockers.bib} 

\end{document}